\documentclass[%
 amsmath,amssymb,
 aps,
prd,showpacs,twocolumn,superscriptaddress
]{revtex4-2}
\usepackage{graphicx}
\usepackage{subfigure}
\usepackage{dcolumn}

\usepackage{amsmath}
\DeclareMathOperator{\arcsinh}{arcsinh}

\usepackage{amssymb}
\usepackage{bm}
\usepackage{color}

\usepackage[normalem]{ulem}
\usepackage[dvipsnames]{xcolor}
\usepackage{hyperref}
\usepackage{bm}

\begin{document}

\preprint{APS/123-QED}

\title{The imprint of cosmic expansion history on the propagation of gravitational waves}

\author{Ye Jiang}
 \affiliation{Shanghai Astronomical Observatory, Shanghai, 200030, China}
\affiliation{School of Astronomy and Space Science, University of Chinese Academy of Sciences,
Beijing, 100049, China}
\author{Wen-Biao Han}%
 \email{wbhan@shao.ac.cn}
\affiliation{Shanghai Astronomical Observatory, Shanghai, 200030, China}
\affiliation{School of Fundamental Physics and Mathematical Sciences, Hangzhou Institute for Advanced Study, UCAS, Hangzhou 310024, China}
\affiliation{School of Astronomy and Space Science, University of Chinese Academy of Sciences,
Beijing, 100049, China}
\affiliation{Taiji Laboratory for Gravitational Wave Universe (Beijing/Hangzhou), University of Chinese Academy of Sciences, Beijing 100049, China;}

\date{\today}

\begin{abstract}
    Gravitational waves (GWs) are regarded as standard sirens for Cosmology. GWs from compact binary coalescence  (CBC) can directly determine the luminosity distance but usually can not obtain information about the redshift. However, if the universe is not flat but accelerating, GWs should carry this cosmological effect. In this Letter, for the first time, we explore how the expansion of the Universe affects GW propagation by perturbing the Robertson-Walker metric. We achieve a comprehensive and rigorous formalism at the next-leading order to describe the cosmological acceleration in GWs from any kind of sources. Theoretically, this cosmological effect will obviously amplify GWs at frequencies as low as $10^{-12}$ Hz.
\end{abstract}

\maketitle


{\it Introduction---}The discovery of gravitational waves (GWs) provides a new means of astronomical observation \cite{PhysRevLett.116.061102}. LIGO \cite{Aasi_2015}, VIRGO \cite{Acernese_2015}, and KAGRA \cite{Somiya_2012} Collaborations (LVK) have observed more than one hundred GW events from comparable mass-ratio compact binaries in their sensitive frequency range ($10\;\rm{Hz}-10^3\;\rm{Hz}$) \cite{PhysRevX.9.031040, PhysRevX.11.021053, PhysRevX.13.041039}. Furthermore, by monitoring an array of millisecond pulsars and measuring the changes in the time-of-arrival of their pulses, the Pulsar Timing Array (PTA) is expected to extend the GW observation in the nano-Hz GW band \cite{Agazie_2023, refId0, Reardon_2023, Xu_2023}. This approach gives us a new perspective on astrophysical problems, early Universe \cite{roshan2024using, PhysRevLett.131.171404, PhysRevLett.127.251303}, dark matter \cite{PhysRevLett.131.171001, PhysRevLett.133.021401, PhysRevResearch.4.L012022}, Hubble constant \cite{PhysRevD.104.063015}, etc.

For the Hubble constant, among novel probes, the standard sirens methodology uses gravitational-wave occurrences to obtain luminosity distances \cite{Abbott2017}. This information is used to infer cosmological parameters, most notably the Hubble constant $H_0$, upon integration with redshift data derived from electromagnetic counterparts or host galaxies. It is disappointing that gravitational waves cannot independently determine the Hubble constant $H_0$. Its theoretical basis is the well-known degeneracy between the chirp mass $M_c$ and redshift $z$, which is directly derived from general relativity together with the flat background; one derives only the redshifted mass $M_c(1+z)$ in the analytical result \cite{PhysRevD.49.2658}.

However, the background is not simply flat, especially when considering a source at cosmological distances. Due to the expansion of the universe, GWs should be regarded as waves propagating in a dynamic space-time. At the same time, past discussions on this problem frequently assumed the flat condition  \cite{PhysRevD.62.083506, PhysRevD.97.104037, PhysRevLett.113.191101} or adopt the geometric optics limit \cite{PhysRevLett.87.221103}, which is inadequate. To deepen our understanding of the propagation of GWs through a dynamic background, we change the background to the Robertson-Walker (RW) metric (including the flat background as a special case), which is one of the most common metrics used to describe a homogeneous, isotropic, expanding universe. Based on this assumption, we derive a new correction factor representing the influence of the expanding universe.

We adopt the cosmological perturbation to solve this problem, which assumes that the entire metric can be separated into two parts $g_{\mu\nu}=\hat{g}_{\mu\nu}+h_{\mu\nu}$, $\hat{g}_{\mu\nu}$ cosmological metric and $h_{\mu\nu}$ represent the gravitational wave. Differing from previous works on cosmological perturbation \cite{PhysRevD.22.1882, PhysRevD.68.103515} involving cosmic evolution or cosmological gravitational wave background, we first apply tensor spherical harmonics to reduce the perturbation to several ordinary differential equations. The main idea was derived from Regge and Wheeler \cite{PhysRev.108.1063}, which have adopted the tensor spherical harmonics to perturb Schwarzschild space-time.

{\it Robertson-Walker metric and Friedmann equation---}
In this section, we will review the background equation of the RW metric. The RW metric describes a homogeneous, isotropic universe that is path-connected and has the following form in reduced-circumference polar coordinates:
\begin{equation}
ds^2=-dt^2+a^2(t)\left(\frac{1}{1-k r^2}dr^2+r^2d\sigma^2\right)
\end{equation}
where $d\sigma^2=d\theta^2+\sin^2\theta d\phi^2$ and the unit is $G=c=1$. In this coordinate, $k\equiv -\Omega_k H_0^2$ denotes the Gaussian curvature of the space at present $t_0$, and $a(t_0) = 1$. $\Omega_k$ and $H_0$ are curvature density and Hubble constant respectively. 

Differing from the Schwarzschild metric, the RW metric is not the vacuum solution of the Einstein field equations; consequently, the perturbation also depends on the formula of energy-moment tensor at first order. By assuming that the energy-moment tensor is homogeneous and isotropic, which means that the nonzero terms of it satisfy $T^{00}=\rho(t)$, $T^{ij}=g^{ij}p(t)$ and solving the Einstein field equations, we obtain the Friedmann equation
\begin{equation}\label{eq: Friedmann equation}
\dot{a}^2+k=\frac{8\pi \rho a^2}{3}.
\end{equation}
Additionally, from the conservation equations $\nabla^{\alpha}T_{\mu\nu}=0$, we have
\begin{equation}\label{eq: conservation equation}
\dot{\rho}+\frac{3\dot{a}}{a}\left(p+\rho\right)=0.
\end{equation}
Now, these equations are sufficient for our following discussion of the perturbation.

{\it RW spacetime perturbation---}We start with the physical setup of our perturbation. As usual, the background manifold of the RW metric should be $\mathbb{S}^3\times\mathbb{R}$ or $\mathbb{R}^4$. However, in our situation, we are not concerned about the detailed structure of the GW source; thus, we remove a spherical region from the background in which the influence of the source is dominant, and the remaining region is still spherically symmetric.

With these physical setups in mind, we then enter our perturbation. The total metric $g_{\mu\nu}$ should satisfy the following Einstein field equation,
\begin{equation}
R_{\mu\nu}=-8\pi S_{\mu\nu},
\end{equation}
where $S_{\mu\nu}=T_{\mu\nu}-1/2g_{\mu\nu}g^{\alpha\beta}T_{\alpha\beta}$.
From the previous discussion, we consider the GW as a small perturbator of spacetime, meaning that the total metric can be written in $g_{\mu\nu}=\hat{g}_{\mu\nu}+h_{\mu\nu}$. Hereafter, $\hat{g}_{\mu\nu}$ denote the RW background and $h_{\mu\nu}$ represent the gravitational wave.

Then, we have the following perturbation equation:
\begin{equation}\label{eq: perturbation equation}
\nabla_\nu{\delta\Gamma^\alpha}_{\mu\alpha}+\nabla_\alpha{\delta\Gamma^\alpha}_{\mu\nu}=-8\pi \delta  S_{\mu\nu},
\end{equation}
where
\begin{equation}
\delta  S_{\mu\nu}=\delta T_{\mu\nu}-\frac{1}{2}\hat{g}_{\mu\nu}\delta {T^\alpha}_\alpha-\frac{1}{2}h_{\mu\nu}{T^\alpha}_\alpha.
\end{equation}
Here, $\nabla_\nu$ denote the covariant derivative corresponding to the background metric $\hat{g}_{\mu\nu}$, and ${\Gamma^\alpha}_{\beta\gamma}$ is the Christoffel symbol related to the total metric $g_{\mu\nu}$.

Note that in our discussion, $\delta T_{\mu\nu}=0$. This is because the GW cannot change the distribution of matter on a large scale. Together with Eq.~(\ref{eq: Friedmann equation}) and Eq.~(\ref{eq: conservation equation}), we have
\begin{equation}
\delta  S_{\mu\nu}=\frac{3}{8\pi}\left(\frac{\Ddot{a}}{a}+\frac{k+\dot{a}^2}{a^2}\right)h_{\mu\nu}.
\end{equation}

Following Regge and Wheeler, we decompose $h$ into tensor functions with the correct symmetry behavior under rotation on the 2-sphere $d\sigma^2$. For the axial perturbation (with the parity $(-)^{l+1}$) for given $l$ and $m$, $h$ has the same form as that of Regge and Wheeler \cite{PhysRev.108.1063}, where the $Y_{lm}(\theta,\phi)$ is the spherical harmonic.

To further simplify the calculation, we selected a suitable gauge for the following discussion. The gauge of the background spacetime is essentially fixed since we have employed the reduced-circumference polar coordinates. However, one can still change the gauge to first order perturbation while keeping the gauge to background. This can be achieved using an infinitesimal gauge transformation ${x'}^{\mu}=x^\mu+\eta^{\mu}$, where the gauge vector $\eta^{\mu}$ is of first order in $h_{\mu\nu}$. This results in a change on $h_{\mu\nu}$ with $\eta_{\mu;\nu}+\eta_{\nu;\mu}$.

For axial perturbations, the gauge vectors should have the following form:
\begin{equation}
\eta^{\mu}=\Lambda(t,r)(0,0,-\frac{1}{\sin\theta}\frac{\partial Y_{lm}}{\partial\phi},\sin\theta\frac{\partial Y_{lm}}{\partial\theta}),
\end{equation}
and takes the change on $h_{\mu\nu}$:
\begin{align}
h'_0=h_0+a^2r^2\frac{\partial\Lambda}{\partial t},\label{eq: gauge h0}\\
h'_1=h_1+a^2r^2\frac{\partial\Lambda}{\partial r},\\
h'_2=h_2-2a^2r^2\Lambda.\label{eq: gauge h2}
\end{align}
It is worth pointing out that $h_0^{\text{inv}}=h_0+(\frac{\partial h_2}{\partial t}-\frac{2\dot{a}h_2}{a})/2$ and $h_1^{\text{inv}}=h_1+(\frac{\partial h_2}{\partial r}-\frac{2h_2}{r})/2$ is invariant under such an infinitesimal coordinate transformation. We will choose the $\Lambda(t,r)=-h_2/(2a^2r^2)$ to make $h_2(t, r)$ become zero and $h_0$, $h_1$ equal to $h_0^{\text{inv}}$, $h_1^{\text{inv}}$, respectively. Under this gauge, only the three components of Eq.~(\ref{eq: perturbation equation}) are independent; therefore, we have three differential equations for only two unknown functions, $h_0$ and $h_1$. After some algebraic simplification, we obtain the differential equation satisfied by $h_1$,
\begin{equation}
\begin{aligned}
&r \left(-r a \dot{a} \frac{\partial h_1}{\partial t}+r a^2 \frac{\partial^2 h_1}{\partial t^2}+k r^3 \frac{\partial^2 h_1}{\partial r^2}\right.\\
&\left.+k r^2 \frac{\partial h_1}{\partial r}-r \frac{\partial^2 h_1}{\partial r^2}+2 \frac{\partial h_1}{\partial r}\right)\\
&+h_1 \left(3 r^2 a \Ddot{a}+7 r^2 \dot{a}^2+3 k r^2+l^2+l-2\right)=0,
\end{aligned}
\end{equation}
and the relationship between $h_0$ and $h_1$,
\begin{equation}\label{eq: h1 to h2}
a \dot{a} h_0+a^2 \frac{\partial h_0}{\partial t}+\left(k r^2-1\right) \frac{\partial h_1}{\partial r}+k r h_1=0,
\end{equation}
Let $h_1(t,r)=\sqrt{a}R(r)T(t)$, one can separate the variable $(t,r)$, and find that
\begin{align}
&\left(3k+\omega^2-\frac{l+l^2-2}{r^2}\right)R-\left(\frac{2}{r}+kr\right)\frac{dR}{dr}\nonumber\\
&+(1-kr^2)\frac{d^2R}{dr^2}=0,\label{eq: h1 radius part}\\
&\left(-\frac{11}{4}\dot{a}^2-\frac{11}{2}a\ddot{a}+\omega ^2\right)T+a^2 \frac{d^2T}{dt^2}=0\label{eq: h1 time part}.
\end{align}

Assuming that $1-kr^2\sim\mathcal{O}(1)$ and dropping the terms $\sim\mathcal{O}(r^{-2})$, the solution to the $h_1$ radial part is
\begin{equation}\label{eq: hypergeo solution}
\begin{aligned}
R(r)&=C_1R_1(r)+C_2R_2(r),\\
R_\pm(r)&=(\frac{r}{\sqrt{1-kr^2}}\pm\frac{i}{\alpha})e^{\pm i\alpha\beta(r)},
\end{aligned}
\end{equation}
where $\alpha=\sqrt{\omega^2+3k}$, $\beta(r)=\int_0^r1/(\sqrt{1-k{r^\prime}^2})dr$ equal to $\arcsin(\sqrt{k}r)/\sqrt{k}$ for $k>0$ or equal to $\arcsinh(\sqrt{-k}r)/\sqrt{-k}$ for $k<0$. Hereafter, we will treat $\alpha$ as $\omega$ due to $k/\omega^2\ll 0$.

For $h_1$ time part, Eq.~(\ref{eq: h1 time part}), by assuming that $\dot{a}\ll\omega$ and $\ddot{a}\ll\omega^2$, we obtain the following asymptotic solution:
\begin{equation}
T(t)=\sqrt{a}(C_1F^+e^{\omega \eta (t)}+C_2F^-e^{-\omega \eta(t)}),
\end{equation}
where $\eta(t)=i\int_{t_\text{emis}}^t1/a(t^\prime)dt^\prime$ and
\begin{equation}\label{eq: amplitude correction}
F^\pm(\omega,t)=1\pm\frac{1}{2i\omega}\int^t_{t_s}\left(3\frac{\dot{a}^2}{a}+5\ddot{a}\right)dt^\prime,
\end{equation}
$t_\text{emis}$ is the GW emission time.
To obtain the explicit form of GW, we change the gauge to the outgoing radiation gauge (ORG), which requires $ah_0-\sqrt{1-kr^2}h_1=0$. Together with Eq.~(\ref{eq: gauge h0}-\ref{eq: gauge h2}) and Eq.~(\ref{eq: h1 to h2}), $\Lambda(t,r)$ should satisfy
\begin{equation}
a^3r^2\Lambda_{,t}-\sqrt{1-kr^2}a^2r^2\Lambda_{,r}=ah_0-\sqrt{1-kr^2}h_1.
\end{equation}

Then, we have
\begin{equation}\label{eq: asy result}
\begin{aligned}
h_+-ih_\times&\sim\frac{1}{a^2r^2}h^{\rm ORG}_2\\
&\sim\frac{F^-(\omega,t)}{a(t)r}\exp\left\{i\omega[\beta(r)-\eta(t)]\right\} \,,
\end{aligned}
\end{equation}
for the outgoing solution of GWs. Where $\omega$ can be regarded as the comoving frequency. Note that we have omitted the $\theta$ and $\phi$ parts of the solution which is ${_{-2}Y}_{lm}$ the spherical harmonic with spin weight $-2$. From our full perturbation scheme, we found that $F^-(\omega,t)$ changes the phase along the geodesic and corrects the GW amplitude at the same time.

If $k=0$ and $a(t)=1$, Eq.~(\ref{eq: asy result}) will become $\exp\left[i\omega (r-t)\right]/r$ which exactly is a monochromatic gravitational wave in flat spacetime.

{\it Discussion---}In this Letter, we study the perturbation of the RW metric and separate variables using spherical tensors. Eq.~(\ref{eq: h1 radius part}) and Eq.~(\ref{eq: h1 time part}) are the radial part and time part, respectively. The solutions to these equations describe the propagation behavior of GWs. From our example for CBC, the correction factor Eq.~(\ref{eq: correction factor}) will become considerable at high redshift and low frequency, e.g., $z>1000$ and $f<10^{-12}\;\text{Hz}$. If one neglects its contribution,  it may result in a non-negligible systemic error in parameter estimation.

To obtain these equations, we have assumed that $\dot{a}^2/\omega^2\rightarrow 0$ and $\ddot{a}/\omega^2\rightarrow 0$. Considering that when redshift moves to infinity, $\dot{a}$ and $\ddot{a}$ are divergence. Our approximate calculation should hold for a finite redshift $z$. However, even if $z$ is as large as 5000 and  $\omega$ as low as $10^{-15}~\text{Hz}$,  $\dot{a}^2/\omega^2\sim 10^{-5}$ and $\ddot{a}/\omega^2\sim 10^{-9}$ are small enough, so our assumption still holds.

In principle, for any spherical GW propagating in RW space-time, there is always a correction factor caused by Eq.~(\ref{eq: asy result}). For example, we now discuss how propagation across cosmological distances modifies the binary inspiral waveform. We assume that there exists a middle radius $r_m$, which means the $r_m$ is sufficiently large so that the gravitational field already has the $1/r$ behavior characteristic of waves but is still sufficiently small so that the expansion of the Universe is negligible. In the region $r<r_m$, $a(t)$ will be a constant, and $k$ can be set as zero. Thus, the waveform can be derived by coordinate transformation $r\rightarrow a_s r$, $a_s=a(t_{\text{emis}})$ is the scale factor at the GW emission time. The result at $r_m$ reads:
\begin{equation}
h(t)=\frac{A}{a_s r_m}{M_{c}}^{5/4}\left(\frac{1}{\tau}\right)^{1/4}\exp [i\Phi(\tau)],
\end{equation}
where $\Phi(\tau)=-2(5M_{c})^{-5/8}\tau^{5/8}+\Phi_0$, $\tau=t_{\text{coal}}-t$, $t_{\text{coal}}$ is the time to coalescence and $\Phi_0=\Phi(\tau=0)$ equal to the value of $\Phi$ at coalescence, $A$ is the amplitude. After performing Fourier transformation and matching the Eq.~(\ref{eq: asy result}), the frequency domain signal can be obtained as follows:
\begin{equation}
h(f)=\frac{A}{L}F^-(f,t_{\text{emis}}){M_{cz}}^{5/6}f^{-7/6}\exp [i\bar{\Phi}(f)]\label{eq: new waveform}
\end{equation}
where $t_c$ and $\phi_c$ are integral constant, $M_{cz}=M_c(1+z)$, $z$ is the redshift at $t_{\text{emis}}$ and $A$ is a constant amplitude parameter. Here, if we only expand $\eta(t)$ to the first order, we can obtain an additional term in phase $\bar{\Phi}$ same with Ref. \cite{PhysRevLett.87.221103} at the flat Lambda cold dark matter  ($\Lambda$-CDM) model. This additional term arises as a result of GWs propagating along the geodesic.

The correction factor $F^-(f,t_{\text{emis}})$ has following form in the $\Lambda$-CDM model:
\begin{align}
&F^-(f,t_{\text{emis}})=1-\frac{H_0}{4if}\int^1_{1/(1+z)}\sigma(x) dx,\label{eq: correction factor}\\
&\sigma(x)=\frac{6 \Omega_k+\Omega_\Lambda(1-15 d )x^{1+3 d}+\Omega_m x-4\Omega_r x^2}{\sqrt{\Omega_r+\Omega_m x+\Omega_k x^2+\Omega_\Lambda x^{-3d}}}.
\end{align}
Here, $\Omega$ are the current day density parameters, with the subscript $m$ for baryons and cold dark matter, $r$ for radiation, $k$ for curvature, and $\Lambda$ for dark energy, $d$ is the parameter of the equation of state of dark energy.

The preview study showed that there exists a degeneracy between the chirp mass $M_c$ and the redshift $z$. This redshift cannot be determined from the detected signal, which is not due to the detecting precision but because the signal is invariant under the transformation $(M_c, L, t)\rightarrow(M_c\lambda, L\lambda, t\lambda)$. One can check that new waveforms Eq.~(\ref{eq: new waveform}) with amplitude correction Eq.~(\ref{eq: amplitude correction}) are not invariant under this transformation, which indicates its potential to break the degeneracy between the chirp mass $M_c$ and the redshift $z$.

There are still some interesting points for future work. In Eq.~(\ref{eq: hypergeo solution}), if $\omega^2+3k<0$, the radial part of $h_1$ will not be an oscillatory solution anymore, but an exponential damping or increase solution instead. This may indicate that a cutoff frequency exists for GWs. Another problem is the polar perturbation, which is a more challenging work. Here, we only consider the axis perturbation, which should be enough to show the basic property of GW propagation. Finally, the LIGO-Virgo-KAGRA and the future space-borne detectors (LISA \cite{amaroseoane2017laser}, Taiji \cite{10.1093/nsr/nwx116} and TianQin \cite{Luo_2016} etc.) are very difficult to observe this cosmological effect. However for GWs with frequency $\lesssim10^{-12}$ Hz, our new correction factor Eq.~(\ref{eq: correction factor}) will effectively amplify the GWs' amplitude.

\nocite{*}

%
\end{document}